\documentclass[usenatbib,usedcolum,usegraphicx]{mn2e}
\usepackage{lscape}
\usepackage{rotating}

\title[
The Catalog of stellar proper motions
in the OGLE-II Galactic bulge fields
      ]{
The Optical Gravitational Lensing Experiment. Catalog of stellar proper motions
in the OGLE-II Galactic bulge fields
 }

\author[               T. ~Sumi, et al. 
       ]
       {             T. ~Sumi$^1$, X. ~Wu$^1$,  A.~Udalski$^2$, M.~Szyma{\'n}ski$^2$, 
        M.~Kubiak$^2$, G.~Pietrzy\'nski$^{2,3}$,  \newauthor 
        I.~Soszy\'nski$^2$, P.~Wo\'zniak$^4$, K.~\.Zebru\'n$^2$,
        O.~Szewczyk$^2$ \& \L.~Wyrzykowski$^2$ \\
    $^1$ Princeton University Observatory, Princeton, NJ 08544-1001, USA;
    e-mail: (sumi, xawn)@astro.princeton.edu \\
$^2$Warsaw University Observatory, Al.~Ujazdowskie~4, 00-478~Warszawa, Poland;\\
~~~e-mail: (udalski,msz,mk,pietrzyn,soszynsk,zebrun,szewczyk,wyrzykow)@astrouw.edu.pl\\
$^3$ Universidad de Concepci{\'o}n, Departamento de Fisica, Casilla 160--C, Concepci{\'o}n, Chile\\
$^4$ Los Alamos National Laboratory, MS-D436, Los Alamos, NM 87545 USA;
e-mail: wozniak@lanl.gov
 \\
}
\date{Accepted 
      Received
      in original form }

\pagerange{\pageref{firstpage}--\pageref{lastpage}}

\begin{document}
\maketitle
\label{firstpage}

\begin{abstract}
We present a proper motion ($\mu$) catalogue of 5,080,236 stars in 
49 Optical Gravitational Lensing Experiment II (OGLE-II) Galactic 
bulge (GB) fields, covering a range of $-11^\circ <l< 11^\circ$ and
$-6^\circ <b<3^\circ$, the total area close to 11 square degrees.
The proper motion measurements are based on 
$ 138 - 555 $ $I$-band images taken during four observing seasons:
1997-2000. The catalogue stars are in the magnitude range
$11 < I < 18$ mag.  In particular, the catalogue includes
Red Clump Giants (RCGs) and Red Giants in the GB,
and main sequence stars in the Galactic disc.
The proper motions up to $\mu = 500 $ mas\,yr $^{-1}$ were
measured with the mean accuracy of 
$0.8 \sim 3.5$ mas\,yr$^{-1}$, depending on the brightness of a star.
This catalogue may be useful for studying the kinematic of stars in 
the GB and the Galactic disk.
\end{abstract}


\section{Introduction}
The Galactic Bulge (GB) is the nearest bulge in which 
individual stars can be studied in detail. A study of 
stellar populations and stellar dynamics in the Bulge may help us
understand how bulges formed, what are their populations, gravitational
potential and structure.

The proper motions with precise photometry might make it possible
to separate the observed populations based on their kinematics.
Such study has first been done by \cite{spa92} with photographic plates
only for a few hundred of brightest red giants in Baade's window.
Recently deeper study has been done by \cite{kui02} with 
{\it Hubble Space Telescope} (HST)/WFPC2 in Baade's window.

Several groups have carried out gravitational microlensing
observations toward dense stellar fields, such as the Magellanic
clouds, the Galactic center and disc.  Until now, hundreds of events
have been found (EROS: \citealt{aub93}; OGLE: \citealt{uda00};
\citealt{woz01}; MACHO: \citealt{alc00}; MOA: 
\citealt{bon01}; \citealt{sum03b}),
and thousands are expected in the upcoming years by MOA
\footnote{{\tt http://www.phys.canterbury.ac.nz/\~{}physib/alert/alert.html}},
OGLE-III
\footnote{ {\tt http://www.astrouw.edu.pl/\~{}ogle/ogle3/ews/ews.html}}
and other collaborations.

It is well known that the gravitational microlensing survey data is
well suited for numerous other scientific projects (see
\citealt{pac96}; \citealt{gou96}).  The studies of the Galactic structure
certainly benefit from this type of data.  The microlensing
optical depth probes the mass density of compact objects along the
line of sight and the event time-scale distribution is related to
the mass function and kinematics of the lensing objects. Observed high
optical depth may be explained by the presence of the bar
(\citealt{uda94}; \citealt{alc97,alc00}; \citealt{sum03b};
 \citealt{afo03}; \citealt{pop03})
There is substantial evidence that the Galaxy has a bar at its center
(\citealt{vau64}; \citealt{bli91}; \citealt{sta94,sta97};
\citealt{kir94}; \citealt{haf00}). However, the parameters of the bar,
e.g., its mass, size, and the motion of stars within it, still remain
poorly constrained.

\cite{sta97} used the Red Clump Giants (RCGs) to constrain the axial ratios
and orientation of the Galactic bar.  These stars are the equivalent of the
horizontal branch stars for a metal-rich population, i.e., relatively
low-mass core helium burning stars.  RCGs in the GB occupy
a distinct region in the colour magnitude diagram (\citealt{sta00} and
references therein).  The intrinsic width of the luminosity distribution of
RCGs in the GB is small, about 0.2 mag 
(\citealt{sta97}; \citealt{pac98}). Their observed peak and width of
the luminosity function are related to the distance and radial depth
of the bar.

Furthermore, \cite{mao02} suggested that there should be a difference in
average proper motions of 1.6 mas\,yr$^{-1}$ between the bright and 
faint RCG sub-samples, which are on average on the near and the far side
of the bar, respectively, if their  tangential streaming motion is
100 km\,s$^{-1}$.  Following this suggestion,
\cite{sum03a} measured mean proper motion of bright and faint RCGs 
in one OGLE-II field in Baade's Window, and they found a difference
to be $ 1.5 \pm 0.11$ mas\,yr$^{-1}$.

To expand this analysis we measured proper motions in all 49 GB
fields observed by the Optical Gravitational Lensing Experiment
\footnote{see {\tt http://www.astrouw.edu.pl/\~{}ogle} or 
{\tt  http://bulge.princeton.edu/\~{}ogle}}
 II (OGLE-II; \citealt{uda00}) for stars down to $I=18$ mag,
which is sufficiently deep to include RCGs.  There are several earlier
proper motion catalogues of this general area (c.f. USNO-B:\citealt{mon02},
Improved NLTT:\citealt{sal03} and Tycho-2:\citealt{hog00}).
Though the area covered by our catalogue is relatively small, it reaches
deeper and covers a wide range of proper motion 
($\mu < 500$ mas\,yr$^{-1}$) with the accuracy 
as good as ($\sim$ 1 mas\,yr$^{-1}$).
In \S\,\ref{sec:data} we describe the data.  We present the analysis 
method in \S\,\ref{sec:analysis} and \ref{sec:HPM}.
In \S\,\ref{sec:catalog}, \ref{sec:zeropoint} and \ref{sec:problem}
we describe the property, zero point and problems in our catalogue.
Discussion and conclusion are given in \S\,\ref{sec:disc}.

\section{DATA}
\label{sec:data}

We use the data collected during the second phase of the
OGLE experiment, between 1997 and 2000. All observations were made
with the 1.3-m Warsaw telescope located at the Las Campanas Observatory,
Chile, which is operated by the Carnegie Institution of Washington.
The "first generation" camera has a SITe 2048 $\times$ 2048 pixel CCD detector
with pixel size of 24 $\mu$m resulting in 0.417 arcsec/pixel scale.
Images of the GB fields were taken in drift-scan mode at "medium" readout
speed with the gain 7.1 $e^{-}$/ADU and readout noise of 6.3 $e^{-}$.
A single 2048 $\times$ 8192 pixel frame covers an area of 0.24 $\times$ 0.95
deg$^2$.  Saturation level is about 55,000 ADU.  Details of the
instrumentation setup can be seen in \cite{uda97}.

In this paper we use 138-555 $I$-band frames of the BUL\_SC1-49 
fields.  The centers of these fields are listed in Table \ref{tbl:fields}.
The time baseline is almost 4 years.  There are gaps between the observing
seasons when the GB cannot be observed from the Earth,
each about 3 months long.  The median seeing is $\sim 1.3''$.
We use the $VI$ photometric maps of the standard OGLE template (\citealt{uda02}) 
as the astrometric and photometric references.

Only about 70\% of the area of the BUL\_SC1 field overlaps with the extinction
map made by \cite{sta96}. The extinction map covering all OGLE-II fields
has been constructed by \cite{sum03c}.

\begin{table*}
\caption{Center positions of OGLE-II 49 fields. The number of frames
$N_{\rm f}$ and stars $N_{\rm s}$, and the mean of uncertainty in proper motion,
$<\sigma_\mu>$ (mas\,yr$^{-1}$), are given as a function
of $I$ (mag) for each field. $\sigma_\mu$ is averaged over 
$ I\pm0.5$ mag.
\label{tbl:fields}}
\begin{tabular}{lcccccrcccccccc}\\
field& $\alpha_{2000}$ & $\delta_{2000}$ & $l$ & $b$ 
  & $N_{\rm f}$ &  $N_{\rm s}$ &
  &  \multicolumn{7}{c}{$\langle \sigma_\mu \rangle$ (mas\,yr$^{-1}$), for $I$\, (mag) =}\\
& & & (deg.) & (deg.) & & 
  & 11.5 & 12.5 & 13.5 & 14.5 & 15.5 & 16.5 & 17.5 \\
\hline
BUL\_SC1& $18^h02^m32.5^s$ & $-29^{\circ}57'41''$ & 1.08&-3.62&259&120697&2.38&0.92&0.88&0.99&1.34&2.53&6.01\\
BUL\_SC2& $18^h04^m28.6^s$ & $-28^{\circ}52'35''$ & 2.23&-3.46&264&140398&2.62&0.94&0.89&1.04&1.40&2.83&6.76\\
BUL\_SC3& $17^h53^m34.4^s$ & $-29^{\circ}57'56''$ & 0.11&-1.93&555&167580&1.98&0.66&0.63&0.78&1.22&2.02&4.66\\
BUL\_SC4& $17^h54^m35.7^s$ & $-29^{\circ}43'41''$ & 0.43&-2.01&514&179906&2.34&0.69&0.69&0.89&1.44&2.50&5.73\\
BUL\_SC5& $17^h50^m21.7^s$ & $-29^{\circ}56'49''$ &-0.23&-1.33&392&113793&1.95&0.78&0.74&0.84&1.12&2.03&3.54\\
BUL\_SC6& $18^h08^m03.7^s$ & $-32^{\circ}07'48''$ &-0.25&-5.70&306&65578&1.78&0.92&0.85&0.94&1.25&2.23&4.47\\
BUL\_SC7& $18^h09^m10.6^s$ & $-32^{\circ}07'40''$ &-0.14&-5.91&323&62357&2.00&0.93&0.90&1.00&1.30&2.27&4.40\\
BUL\_SC8& $18^h23^m06.2^s$ & $-21^{\circ}47'53''$ &10.48&-3.78&289&52549&2.05&0.90&0.86&0.95&1.20&1.99&3.79\\
BUL\_SC9& $18^h24^m02.5^s$ & $-21^{\circ}47'55''$ &10.59&-3.98&288&51274&1.82&0.90&0.85&0.91&1.17&1.96&3.71\\
BUL\_SC10& $18^h20^m06.6^s$ & $-22^{\circ}23'03''$ & 9.64&-3.44&291&57064&1.80&0.99&0.93&0.97&1.27&2.13&4.09\\
BUL\_SC11& $18^h21^m06.5^s$ & $-22^{\circ}23'05''$ & 9.74&-3.64&280&51181&1.87&1.01&0.93&0.96&1.24&2.01&3.83\\
BUL\_SC12& $18^h16^m06.3^s$ & $-23^{\circ}57'54''$ & 7.80&-3.37&294&79162&2.26&0.98&0.93&1.02&1.36&2.37&4.71\\
BUL\_SC13& $18^h17^m02.6^s$ & $-23^{\circ}57'44''$ & 7.91&-3.58&270&79082&2.21&1.04&1.02&1.07&1.47&2.50&4.93\\
BUL\_SC14& $17^h47^m02.7^s$ & $-23^{\circ}07'30''$ & 5.23& 2.81&290&90091&2.39&1.01&0.97&1.05&1.44&2.41&5.23\\
BUL\_SC15& $17^h48^m06.9^s$ & $-23^{\circ}06'09''$ & 5.38& 2.63&285&84372&2.19&0.98&0.94&1.03&1.39&2.26&4.88\\
BUL\_SC16& $18^h10^m06.7^s$ & $-26^{\circ}18'05''$ & 5.10&-3.29&277&100885&2.30&1.01&0.93&1.03&1.42&2.57&5.49\\
BUL\_SC17& $18^h11^m03.6^s$ & $-26^{\circ}12'35''$ & 5.28&-3.45&284&101955&2.21&0.95&0.92&1.02&1.40&2.58&5.50\\
BUL\_SC18& $18^h07^m03.5^s$ & $-27^{\circ}12'48''$ & 3.97&-3.14&275&133282&2.48&0.97&0.91&1.08&1.47&2.91&6.64\\
BUL\_SC19& $18^h08^m02.4^s$ & $-27^{\circ}12'45''$ & 4.08&-3.35&273&112421&2.39&0.97&0.93&1.02&1.38&2.53&5.84\\
BUL\_SC20& $17^h59^m19.1^s$ & $-28^{\circ}52'55''$ & 1.68&-2.47&316&169423&2.70&0.99&0.98&1.15&1.67&3.36&8.10\\
BUL\_SC21& $18^h00^m22.3^s$ & $-28^{\circ}51'45''$ & 1.80&-2.66&321&161268&2.64&0.99&0.96&1.13&1.63&3.34&7.77\\
BUL\_SC22& $17^h56^m47.6^s$ & $-30^{\circ}47'46''$ &-0.26&-2.95&414&111768&2.32&0.79&0.77&0.83&1.19&1.92&4.59\\
BUL\_SC23& $17^h57^m54.5^s$ & $-31^{\circ}12'36''$ &-0.50&-3.36&350&94798&2.15&0.80&0.73&0.79&1.10&1.74&4.10\\
BUL\_SC24& $17^h53^m17.9^s$ & $-32^{\circ}52'45''$ &-2.44&-3.36&359&91733&2.20&0.87&0.80&0.85&1.19&1.88&4.22\\
BUL\_SC25& $17^h54^m26.1^s$ & $-32^{\circ}52'49''$ &-2.32&-3.56&342&90853&2.48&0.84&0.78&0.86&1.14&1.85&4.30\\
BUL\_SC26& $17^h47^m15.5^s$ & $-34^{\circ}59'31''$ &-4.90&-3.37&346&95233&2.00&0.85&0.81&0.94&1.31&2.28&5.03\\
BUL\_SC27& $17^h48^m23.6^s$ & $-35^{\circ}09'32''$ &-4.92&-3.65&334&92508&2.11&0.88&0.84&0.94&1.29&2.22&4.90\\
BUL\_SC28& $17^h47^m05.8^s$ & $-37^{\circ}07'47''$ &-6.76&-4.42&321&57501&2.15&0.83&0.75&0.79&0.98&1.64&3.43\\
BUL\_SC29& $17^h48^m10.8^s$ & $-37^{\circ}07'21''$ &-6.64&-4.62&313&57822&2.38&0.81&0.76&0.78&1.00&1.66&3.41\\
BUL\_SC30& $18^h01^m25.0^s$ & $-28^{\circ}49'55''$ & 1.94&-2.84&323&151877&2.65&0.91&0.87&1.05&1.51&2.94&7.05\\
BUL\_SC31& $18^h02^m22.6^s$ & $-28^{\circ}37'21''$ & 2.23&-2.94&334&155383&2.22&0.95&0.92&1.11&1.54&3.11&7.42\\
BUL\_SC32& $18^h03^m26.8^s$ & $-28^{\circ}38'02''$ & 2.34&-3.14&313&155100&2.31&0.93&0.91&1.10&1.55&3.21&7.56\\
BUL\_SC33& $18^h05^m30.9^s$ & $-28^{\circ}52'50''$ & 2.35&-3.66&273&121848&2.57&0.92&0.85&1.00&1.34&2.56&6.01\\
BUL\_SC34& $17^h58^m18.5^s$ & $-29^{\circ}07'50''$ & 1.35&-2.40&329&156281&2.49&0.99&0.95&1.16&1.70&3.04&7.54\\
BUL\_SC35& $18^h04^m28.6^s$ & $-27^{\circ}56'56''$ & 3.05&-3.00&260&139324&2.51&0.96&0.89&1.05&1.45&2.87&6.80\\
BUL\_SC36& $18^h05^m31.2^s$ & $-27^{\circ}56'44''$ & 3.16&-3.20&290&145376&2.70&0.99&0.95&1.10&1.55&3.27&7.92\\
BUL\_SC37& $17^h52^m32.2^s$ & $-29^{\circ}57'44''$ & 0.00&-1.74&406&152704&2.46&0.75&0.72&0.87&1.27&2.13&4.49\\
BUL\_SC38& $18^h01^m28.0^s$ & $-29^{\circ}57'01''$ & 0.97&-3.42&268&123428&2.40&0.93&0.90&1.05&1.43&2.66&6.41\\
BUL\_SC39& $17^h55^m39.1^s$ & $-29^{\circ}44'52''$ & 0.53&-2.21&415&155735&2.46&0.79&0.77&0.91&1.38&2.26&5.53\\
BUL\_SC40& $17^h51^m06.1^s$ & $-33^{\circ}15'11''$ &-2.99&-3.14&325&82152&2.11&0.89&0.82&0.88&1.19&1.85&3.99\\
BUL\_SC41& $17^h52^m07.2^s$ & $-33^{\circ}07'41''$ &-2.78&-3.27&312&87013&2.44&0.90&0.86&0.91&1.24&1.96&4.34\\
BUL\_SC42& $18^h09^m05.0^s$ & $-26^{\circ}51'53''$ & 4.48&-3.38&273&99152&2.40&0.98&0.94&1.03&1.45&2.50&5.31\\
BUL\_SC43& $17^h35^m13.5^s$ & $-27^{\circ}11'00''$ & 0.37& 2.95&382&76840&1.86&0.80&0.82&0.84&1.14&1.91&3.65\\
BUL\_SC44& $17^h49^m22.4^s$ & $-30^{\circ}02'45''$ &-0.43&-1.19&343&68457&1.85&0.86&0.86&0.90&1.17&1.98&3.75\\
BUL\_SC45& $18^h03^m36.5^s$ & $-30^{\circ}05'00''$ & 0.98&-3.94&140&107362&3.13&1.67&1.65&1.85&2.45&4.62&10.57\\
BUL\_SC46& $18^h04^m39.7^s$ & $-30^{\circ}05'11''$ & 1.09&-4.14&138&97197&3.38&1.86&1.78&1.94&2.46&4.45&10.43\\
BUL\_SC47& $17^h27^m03.7^s$ & $-39^{\circ}47'16''$ &-11.19&-2.60&242&47459&3.01&1.32&1.29&1.33&1.59&2.47&4.74\\
BUL\_SC48& $17^h28^m14.0^s$ & $-39^{\circ}46'58''$ &-11.07&-2.78&237&47673&2.65&1.39&1.29&1.28&1.56&2.44&4.70\\
BUL\_SC49& $17^h29^m25.1^s$ & $-40^{\circ}16'21''$ &-11.36&-3.25&234&43341&3.09&1.29&1.26&1.25&1.50&2.32&4.51\\
\end{tabular}
\end{table*}

\section{Analysis}
\label{sec:analysis}

The analysis in this work follows \cite{sum03a}, except our procedure
makes it possible to detect high proper motions, extends to the
limiting magnitude down to $I=18$ mag and corrects the systematic effects. 
The standard OGLE template given by \cite{uda02} serves 
as the fixed astrometric reference in our analysis.
In order to treat properly frame distortions in the y-axis (declination) 
due to drift-scan mode of observation 
each OGLE-II field is divided into 64 subframes before processing. 
Subframes are 2048$\times$128 pixels with a 14 pixel margin on each side.

We compute the pixel positions of stars in the images 
using the DoPHOT package (\citealt{sch93}).  At the start of the
processing for each exposure, the positions of stars
in a single subframe are measured and cross-referenced with those in
the template and the overall frame shift is obtained.  Using this
crude shift we can identify the same region of the sky (corresponding
to a given subframe of the template) throughout the entire sequence 
of frames.

To treat properly spatial PSF variations, each 2048$\times$128 pixel
subframe is divided into 4 smaller chunks with a size of 
512$\times$128 pixels with a 14 pixel margin on each side.
Then the positions of stars in all chunks are computed by DoPHOT.
We use all stars with $I\le18$ mag ($\sim 400$ of them, depending on 
the stellar density in each field) categorized by DoPHOT as isolated 
stars (marked as type=1) in the following analysis.  We do not use the 
data points categorized by DoPHOT as a star blended with other stars
(marked as type=3).

The stars in each of the chunks are combined into the original 
2048$\times$128 subframes.
We cross-reference the stars in the template and other frames
with a search radius of 0.5 pixels and derive the local 
transformation between these pixel coordinate systems for each subframe.
We use a first order polynomial to fit the transformation.  
The resulting piece-wise transformation adequately
converts pixel positions to the reference frame of the template. Typical
residuals are at the level of $0.08$ pixels for bright stars ($I<16$) and
$0.2$ pixels for all stars ($I<18$).

By using these transformation matrices, we cross-reference the 
stars in the template and other frames with a search radius of
1.0 pixels instead of 0.5 pixels used in \cite{sum03a} to increase 
the range of detectable high proper motion objects. 
We estimate that the probability of the mis-identification in 
this search radius is negligible (0.26 \%).

\begin{figure}
\begin{center}
\includegraphics[angle=-90,scale=0.37,keepaspectratio]{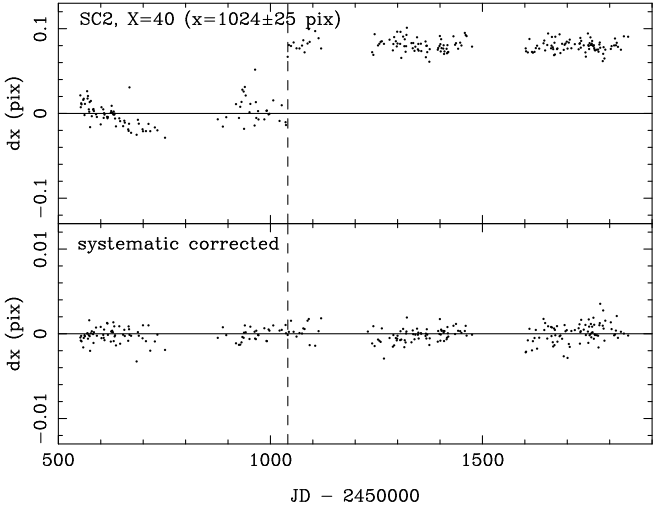}
\caption{Upper panel: The mean positional shift in $x$, 
$\langle dx \rangle$ of stars in a strip 
$X=40$ ($x=1024\pm25$ pixels) in BUL\_SC2
as a function of time. We can see the big jump at 
JD=2451041 (indicated by a vertical dashed line) where the exposure 
time of OGLE-II in the GB fields has changed from 87 sec to 99 
sec in the middle of 1998 season, on August 15.
Lower panel: The same figure as above after the systematic correction.
  \label{fig:dxvst}
  }
\end{center}
\end{figure}

\begin{figure}
\begin{center}
\includegraphics[angle=-90,scale=0.37,keepaspectratio]{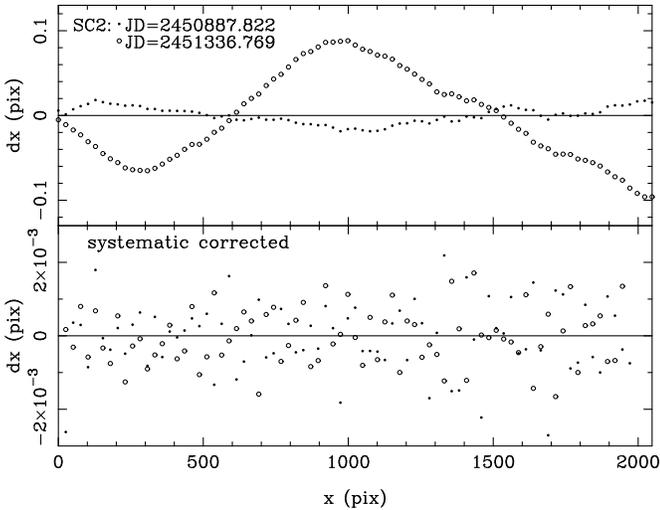}
\caption{Upper panel: Typical mean positional shifts in $x$, 
$\langle dx \rangle$ of stars in strips in BUL\_SC2 as a function of 
pixel coordinate $x$. The filled and open circles represent the 
$\langle dx \rangle$ of the frame taken at 
JD=2450887.822 (before the jump) and 2451336.769 (after the jump), 
respectively.
Lower panel: The same figure as above after the systematic correction.
Note that vertical scale in the lower panel is very different than in
the upper panel.
  \label{fig:dxvsx}
  }
\end{center}
\end{figure}

We have found that there are systematic differences in 
the mean positional shifts of stars from the template position 
$\langle dx \rangle$ and $\langle dy \rangle$  depending on time and 
pixel coordinate in $x$.
We have measured the $\langle dx \rangle$ and $\langle dy \rangle$ 
of the stars in 81 strips ($X=0\sim 80$) centered at equal intervals
in x coordinate between $0\le x \le 2048$ with a width of  $\pm25$ 
pixels.  Each strip contains typically $\sim 2,000$ stars.
In the upper panel of Fig. \ref{fig:dxvst} we show $\langle dx \rangle$
as a function of time for the strip $X=40$ ($x=1024\pm25$ pixels) in 
BUL\_SC2.  We can see the big jump at 
JD=2451041 (indicated by a vertical dashed line) where the exposure 
time of OGLE-II in the GB fields has been changed from 87 sec to 99 
sec in the middle of 1998 season, on August 15.
In the upper panel of Fig. \ref{fig:dxvsx} we show typical mean 
positional shifts in $x$, $\langle dx \rangle$ of stars in strips 
in BUL\_SC2 as a function of pixel coordinate $x$. The filled and 
open circles represent the $\langle dx \rangle$ of the frame taken
at 
JD=2450887.822 (before the jump) and 2451336.769 (after the jump),  
respectively.
There are also systematics in  $\langle dy \rangle$ with the level
of 0.04 pixels. We cannot see any such systematics as a function
of $y$ pixel coordinate.
Because of the good coincidence between the jump and the change in the
drift scan rate that determines the effective exposure time,
the bulk of the systematics may be caused by the change in
drift scan rate. However, we do not know the detail reasons behind this 
at the present time.

Even within the period before and after the jump, the shapes of 
Fig. \ref{fig:dxvsx}  differ from time to time and from field 
to field at the level of 0.04 pixels.
By interpolating these curves of $\langle dx \rangle$ and 
$\langle dy \rangle$ as a function of $x$ for each frame (time) 
of each field, we correct $dx$ and $dy$ for each star and frame.
In the lower panel of Fig. \ref{fig:dxvst} and Fig. \ref{fig:dxvsx},
we show the same plots after this systematic correction.
This procedure is based on the assumption that average proper motions
of a large number of stars in separate groups of columns (i.e. 
different values of X) should be the same. Note that the integral
of the curves shown in Fig. \ref{fig:dxvsx} over all x-columns is
unity, as this corresponds to the average position of all stars.

\begin{figure}
\begin{center}
\includegraphics[angle=-90,scale=0.37,keepaspectratio]{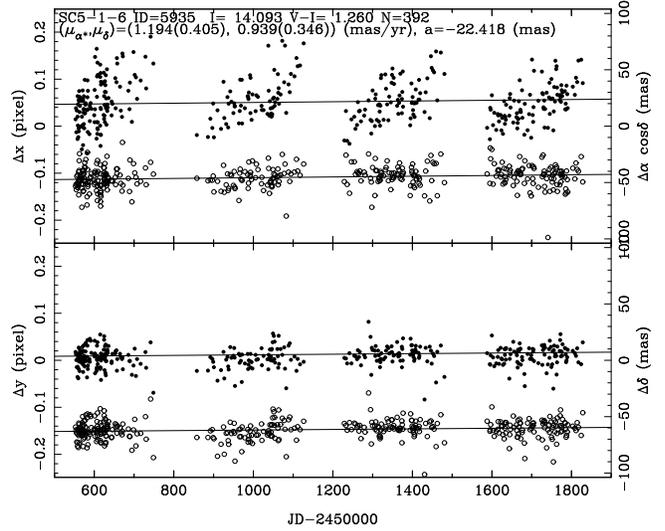}
\caption{Time variation of the position in $\alpha{\rm cos}\delta$
(upper) and $\delta$ (lower) for star ID=5935 (V-I=1.260) in BUL\_SC5.
Filled circles represent actual positions, while open circles are positions
corrected for differential refraction with an offset of $-0.16$ pixels.
Solid lines indicate a model fit for the proper motion
$(\mu_{\alpha*}, \mu_\delta)=(-1.19 (0.41), 0.94 (0.35))$ (mas\,yr$^{-1}$) with the 
1 $\sigma$ errors in bracket. The differential
refraction coefficient is $a=-22.42$ mas.
  \label{fig:shiftcurve}
  }
\end{center}
\end{figure}

An example of time dependence of the position for a star with a moderate
detectable proper motion is shown in Fig. \ref{fig:shiftcurve} with filled
circles.  To measure the proper motions in right ascension 
($\mu_{\alpha*}\equiv \mu_\alpha cos\delta$) and in declination 
($\mu_\delta$), we fit the positions as a function of time $t$ with 
the following formula:

\begin{equation}
  \label{eq:pm_model_a}
   \alpha = \alpha_0 + \mu_{\alpha*} t + a\sin C\tan z, 
\end{equation}

\begin{equation}
  \label{eq:pm_model_d}
   \delta = \delta_0 + \mu_\delta t + a\cos C\tan z, 
\end{equation}
where $a$ is the coefficient of differential refraction, $z$ is the
zenith angle, and $C$ is the angle between the line joining the star and
Zenith and the line joining the star with the South Pole, and
$\alpha_0$ and $\delta_0$ are constants. The parameter $a$
is a function of the apparent star colour.  We neglect the parallactic
motion due to the Earth's orbit because its effect is strongly
degenerate with the effect of differential refraction for stars in the
direction of the GB (\citealt{eye01}).

In Fig. \ref{fig:shiftcurve} we present with solid lines and open circles 
the best fit model for the proper motion $(\mu_{\alpha*}, \mu_\delta)$ and 
positions allowing for the differential refraction respectively. As is
written in this figure, the parameters for this object are: field SC5-1-6, 
which means that this object is in the OGLE-II field BUL\_SC5, and chunk 
($X_{chunk}, Y_{chunk}$)=(1, 6), OGLE ID=5935, $I=14.093$, $V-I=1.260$, 
the number of data 
points $N=392$, proper motion $(\mu_{\alpha*}, \mu_\delta)=
(1.19 (0.41), 0.94 (0.35))$ (mas\,yr$^{-1}$) 
with 1 $\sigma$ errors in brackets. The differential
refraction coefficient is $a=-22.42$ mas.

We computed $\alpha_0$, $\delta_0$, $\mu_{\alpha*}$, $\mu_\delta$ and $a$
for all stars used to transform coordinate systems (approximately
the number of fields times the number of chunks times the typical 
number of stars per chunk, i.e. 49 $\times$ 256 $\times$ 400). In cases 
where the star is measured in the overlap region of more than one chunk 
of a given field, the data set with the largest number of points is
selected.  Stars with the fewer than 20 data points are rejected.
The catalogue of whole BUL\_SC1 was recomputed in this analysis because 
the catalogue presented by \cite{sum03a} contains only 70 \% of this field.

Our catalogue contains 5,080,236 stars, which is 79.8 \% of all objects
 with $I\le18$ mag in the original OGLE template (\citealt{uda02}).
7\% of these stars do not have any $V$-band photmetry.
The missing $V$-band photometry of stars can be estimated from the 
differential refraction coefficient because there is good correlation
between the differential refraction coefficient $a$ and the apparent 
$V-I$ colour for stars (see Fig. \ref{fig:difref}), provided they belong to the 
same population as that of the majority.

\begin{figure}
\begin{center}
\includegraphics[angle=0,scale=0.4,keepaspectratio]{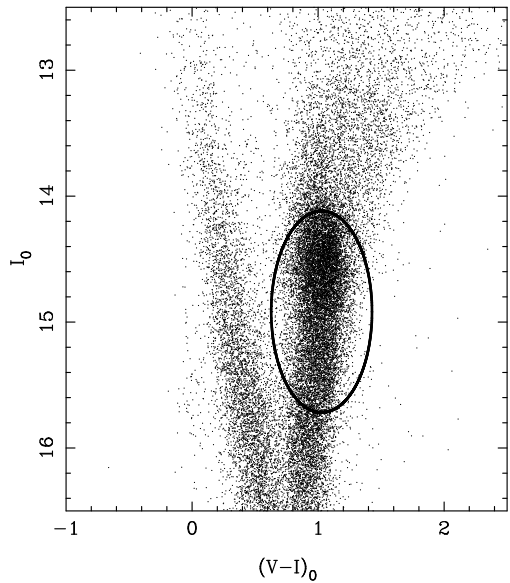}
\caption{ Colour Magnitude Diagram of stars with $\sigma_\mu<2.5$ mas\,yr$^{-1}$
 in BUL\_SC2.  $I_0$ and $(V-I)_0$ are extinction corrected $I$-band magnitude 
and $V-I$ colour. The stars in GB are defined within the ellipse
centered at the center of RCGs plus 0.4 mag in $I$.
  \label{fig:cmd}
  }
\end{center}
\end{figure}

\begin{figure}
\begin{center}
\includegraphics[angle=-90,scale=0.37,keepaspectratio]{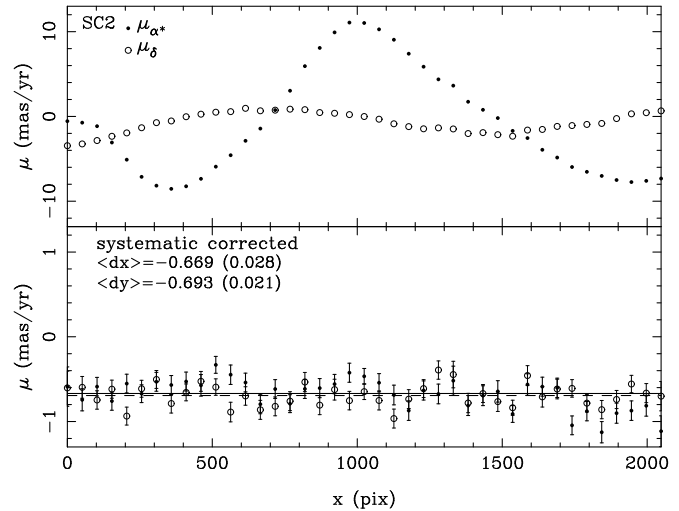}
\caption{Upper panel: The mean proper motions in $x$ (filled circle) and
in $y$ (open circle), for stars in 50 pixels strips in BUL\_SC2 as a function
of pixel coordinate $x$. 
Lower panel: The same figure as above after the systematic correction.
  \label{fig:muvsx}
  }
\end{center}
\end{figure}

We have measured the mean proper motions of stars in the Galactic bulge
(GB) defined in the ellipse in the Colour Magnitude Diagram (CMD) in Fig. 
\ref{fig:cmd}, where the extinction and reddening are corrected by using the 
extinction map of \cite{sum03c}.  The ellipse is located at
the center of the RCGs estimated in \cite{sum03c} plus 0.4 mag in $I$ and have 
semi-major and minor axises of 0.9 mag and 0.4 mag, respectively.
This ellipse includes RCGs and Red Giants in the GB.
Here we chose only objects whose proper motion accuracy is better than 
2.5 mas\,yr$^{-1}$.  These mean proper 
motions of stars in the GB are assumed to be constant within each field. 
In Fig. \ref{fig:muvsx} we show these measured mean proper 
motions of stars in the GB as a function of $x$ (upper panel) and after 
the systematic correction (lower panel). 
We can see how well the systematic distortions are corrected.
Note that our instrumental reference frame is defined by all stars, 
and this is why the average proper motion of stars in the GB are not zero.

\section{High proper motion objects}
\label{sec:HPM}

The detectability of high proper motions is limited by the search 
radius used for cross -- identification of stars on all images. 
The search radius, 1 pix\,yr$^{-1}$, corresponds to $ \sim 400$ 
mas\,yr$^{-1}$.
Objects with $\mu \ge 100$ mas\,yr$^{-1}$ cannot be identified
over the full four year long observing interval of OGLE-II, as they move
out of the search radius.  
In order to be able to follow fast moving stars over four observing
seasons we made additional astrometry for objects for which preliminary
estimates gave a proper motion $ \mu \ge 100 $ mas\,yr$^{-1}$.  
Whenever a star moved more than 0.4 pixels from the previous search center
we moved that center to the median location of the last 3 data points.
This procedure was adopted when it allowed us to locate the star in a
larger number of CCD images.

\begin{figure}
\begin{center}
\includegraphics[angle=-90,scale=0.37,keepaspectratio]{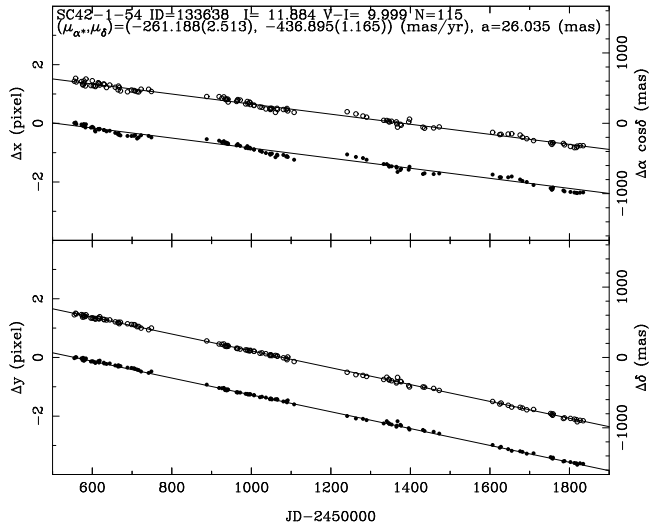}
\caption{Same plots as Fig. \ref{fig:shiftcurve} for
ID=133638 (V-I=none) in OGLE-II field BUL\_SC42, one of the
highest proper motion stars. 
Filled circles represent the actual positions and open circles are the positions
corrected for differential refraction, with an offset of $+1.5$ pixels.
Solid lines indicate a model fit for the proper motion $(\mu_{\alpha*},
\mu_\delta)=(-261.19 (2.51), -436.90 (1.2))$ (mas\,yr$^{-1}$) with
1 $\sigma$ errors in bracket. The differential
refraction coefficient is $a=26.04$ mas, which indicates the star is very red.
This star dose not have $V$-band magnitude because of the high proper motion.
The photometries of the star obtained by hand relatively to the neighboring stars
are: $I=11.70$, $V-I=2.86$. 
  \label{fig:shiftcurveHPM}
  }
\end{center}
\end{figure}

\begin{figure}
\begin{center}
\includegraphics[angle=0,scale=0.45,keepaspectratio]{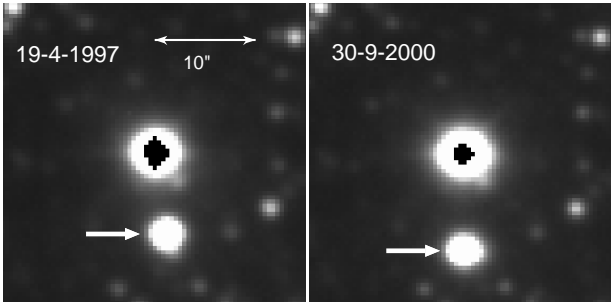}
\caption{Images of the high proper motion star in Fig. 
\ref{fig:shiftcurveHPM}, i.e. ID=133638 in OGLE-II field BUL\_SC42, 
at 19 April 1997 (left) and at 30 September 2000 (right).
  \label{fig:imHPM}
  }
\end{center}
\end{figure}

We show the positional movement of one of the highest proper motion
objects in Fig. \ref{fig:shiftcurveHPM} and images at 1997 and at 2000
in Fig. \ref{fig:imHPM}. This star has relatively
fewer data points because this star is overexposed on the good 
seeing frames in the $I$-band.  This star dose not have any $V$-band
measurements because of the failure of cross-identification in the 
$V$-band template image due to its high proper 
motion. $I$-band measurements of such objects may also be unreliable 
because the OGLE photometric maps are based on the measurements by the 
"fixed position mode" over years (\citealt{uda02}).  The photometries of 
this star obtained by hand relative to the neighboring stars are: 
$I=11.70$, $V-I=2.86$. This colour is very red as expected from the 
relation between differential refraction coefficient $a$ from the fit 
and colour in Fig.\ref{fig:difref}, but its $a$ and $V-I$ do not match
 the relation exactly. 

The relation between $a$ and $V-I$ for BUL\_SC42 in which the extinctions
are relatively small, is slightly different from 
that for BUL\_SC5 plotted in Fig. \ref{fig:difref}.  The zero-point 
of $a$ is 15 mas higher than that for BUL\_SC5, but the expected 
colour of this star from this relation is still redder ($V-I=4-5$).
In these figures the slopes differ between fields for the region $V-I >3$.
This is because of the difference of the population.  In  BUL\_SC5, the 
majority in this colour region are RCGs and Red Giant Branch stars,
but Red Super Giants in  BUL\_SC42.  So this disagreement might be 
because this object is a nearby very red dwarf of M4-5 spectral type, 
not typical for this catalogue.

\section{Catalogue}
\label{sec:catalog}

A sample for our catalogue of proper motions is shown in Table
\ref{tbl:list}.  The complete list of all  5,080,236 stars is 
available in electronic format via anonymous ftp from the server 
{\it ftp://ftp.astrouw.edu.pl/ogle/ogle2/proper\_motion/} and
{\it ftp://bulge.princeton.edu/ogle/ogle2/proper\_motion/}.
The list contains star ID, the number 
$\mu_{\alpha*}$ and $\mu_\delta$ and in galactic coordinate 
$\mu_{l}$ and $\mu_b$ with their errors, differential refraction
coefficient $a$, standard deviation (Sdev) of data points in the
fitting, equatorial coordinates $\alpha_{2000}$, $\delta_{2000}$ in 2000, 
apparent $I$-band magnitude and $V-I$ colour, pixel coordinates on 
CCD $x$ and $y$, and position of chunk $X_c$ and $Y_c$ to which this
object belongs.  ID, 2000 coordinates, $I$ and $V-I$ for each 
object are identical to those in \cite{uda02}.
If the object doesn't have $V$-band photometry the $V-I$ colour 
is written as 9.999.

Note that the number of data points $N$ differs from star to star even if 
they have similar brightness. This is because some are near the edge of
the CCD image or CCD defects, and others are affected by blending. We use
the positions which are categorized as a single isolated star by DoPHOT.
Hence, blended stars can not be measured when the seeing is poor.

\begin{table}
\rotcaption{Sample of proper motion catalogue for BUL\_SC2.
}
\begin{sideways}
    \label{tbl:list}
    \begin{tabular}{lcrcrcrrcccccccccccc}\\
  ID &N& 
 $\mu_{\alpha*}$ &$\sigma_{\mu_{\alpha*}}$& $\mu_\delta$ & $\sigma_{\mu_\delta}$ 
&$\mu_l$ &$\sigma_{\mu_l}$& $\mu_b$ & $\sigma_{\mu_b}$ 
& $a$  & Sdev & $\alpha_{2000}$ & $\delta_{2000}$ & $I$ & $V-I$ 
& $x$  & $y$ & $X_c$ & $Y_c$ \\
     && \multicolumn{8}{c}{(mas\,yr$^{-1}$)} & \multicolumn{2}{c}{(mas)} &
      (deg)&(deg.)& \multicolumn{2}{c}{(mag)} & \multicolumn{2}{c}{(pixel)} \\
  \hline
13213&203& 2.96&0.90& -0.44&0.63  & 1.06&0.70& -2.80&0.84    &  0.69& 13.02 &270.99213&-29.24292&15.282&1.522 & 62.52&904.72&1&7\\
13214&258& 1.09&0.66& -2.03&0.41  &-1.24&0.48& -1.94&0.61    & -0.88& 10.16 &271.03837&-29.24297&15.507&1.716 &413.60&904.92&1&8\\
13215&255&-1.71&0.68& -4.75&1.01  &-4.98&0.94& -0.82&0.77    &  1.18& 16.10 &271.04729&-29.24297&15.572&1.599 &481.45&904.94&1&7\\
13216&247&-0.11&0.67& -1.46&0.51  &-1.33&0.55& -0.62&0.64    & -0.57& 10.95 &271.00104&-29.24281&15.635&1.990 &130.10&905.70&1&7\\
13217&250&-3.44&0.83& -0.74&0.76  &-2.32&0.78&  2.64&0.81    & -2.72& 14.72 &271.00654&-29.24286&15.768&1.746 &171.88&905.41&1&7\\
13218&264& 1.49&1.01& -0.30&0.66  & 0.46&0.76& -1.45&0.94    & -3.97& 16.19 &271.04246&-29.24286&14.960&1.711 &444.59&905.97&1&8\\
13219&263&-3.31&0.91& -1.75&0.39  &-3.14&0.56&  2.04&0.82    & -2.80& 13.20 &271.02271&-29.24231&15.132&1.931 &294.82&910.42&1&8\\
13220&247&-1.11&0.78& -4.67&1.41  &-4.62&1.29& -1.31&0.97    &  2.30& 20.88 &270.99650&-29.24206&15.436&1.644 & 95.64&912.22&1&8\\
13222&262&-3.98&0.55& -6.50&0.50  &-7.62&0.51&  0.31&0.54    &  0.29&  9.87 &271.01058&-29.24183&15.310&1.853 &202.60&914.42&1&8\\
13223&245& 2.25&1.11& -7.76&1.14  &-5.68&1.13& -5.75&1.12    &  5.67& 20.22 &271.04275&-29.24197&15.369&1.903 &446.73&913.63&1&8\\
13224&189&-2.72&0.87&  0.31&0.64  &-1.06&0.70&  2.53&0.82    & -3.43& 12.55 &270.99100&-29.24144&15.874&1.915 & 53.69&917.54&1&8\\
13225&131& 0.83&0.88& -3.41&1.01  &-2.57&0.98& -2.39&0.91    &  1.48& 13.38 &270.98617&-29.24125&15.460&1.795 & 17.09&919.06&1&8\\
13227&248& 0.32&0.53& -5.23&0.68  &-4.41&0.65& -2.83&0.57    &  0.41& 11.08 &271.03579&-29.24092&15.015&1.690 &394.10&922.69&1&8\\
13228&262& 0.45&0.60&  1.12&0.62  & 1.20&0.62&  0.15&0.61    &  0.31& 11.51 &271.00475&-29.24047&15.208&1.909 &158.16&926.03&1&8\\
13229&245&-1.94&0.92& -6.90&0.98  &-6.97&0.97& -1.67&0.93    &  0.10& 17.20 &270.99608&-29.24036&15.402&1.876 & 92.42&926.87&1&8\\
13230&264& 3.89&0.57&  3.85&0.60  & 5.26&0.59& -1.52&0.58    & -0.14& 11.09 &271.02446&-29.24042&15.082&1.757 &307.89&926.93&1&8\\
13231&249& 0.82&0.76& -0.58&0.68  &-0.11&0.70& -1.00&0.74    & -1.17& 13.26 &271.03533&-29.24047&15.090&1.850 &390.64&926.65&1&8\\
13232&200& 2.53&1.72& 11.04&1.50  &10.87&1.56&  3.17&1.67    &  0.86& 26.09 &271.00042&-29.24025&15.608&1.658 &125.33&928.07&1&8\\
\end{tabular}
\end{sideways}
\end{table}

\begin{figure}
\begin{center}
\includegraphics[angle=0,scale=0.4,keepaspectratio]{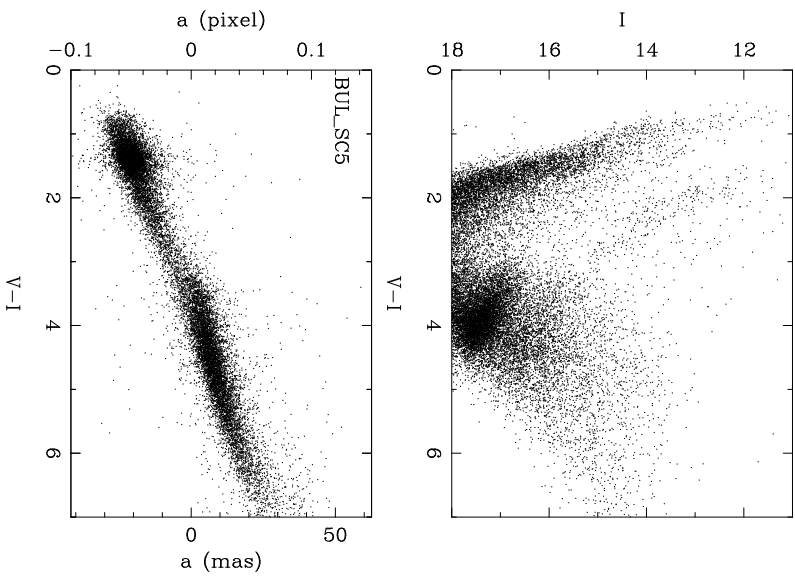}
\caption{Upper: Colour -- magnitude diagram of a quarter of the analyzed 
stars in OGLE-II field BUL\_SC5.
Lower: correlation between the differential refraction
coefficient $a$ and the apparent colour for stars with $I<16$ in this field.
  \label{fig:difref}
  }
\end{center}
\end{figure}
\begin{figure}
\begin{center}
\includegraphics[angle=0,scale=0.45,keepaspectratio]{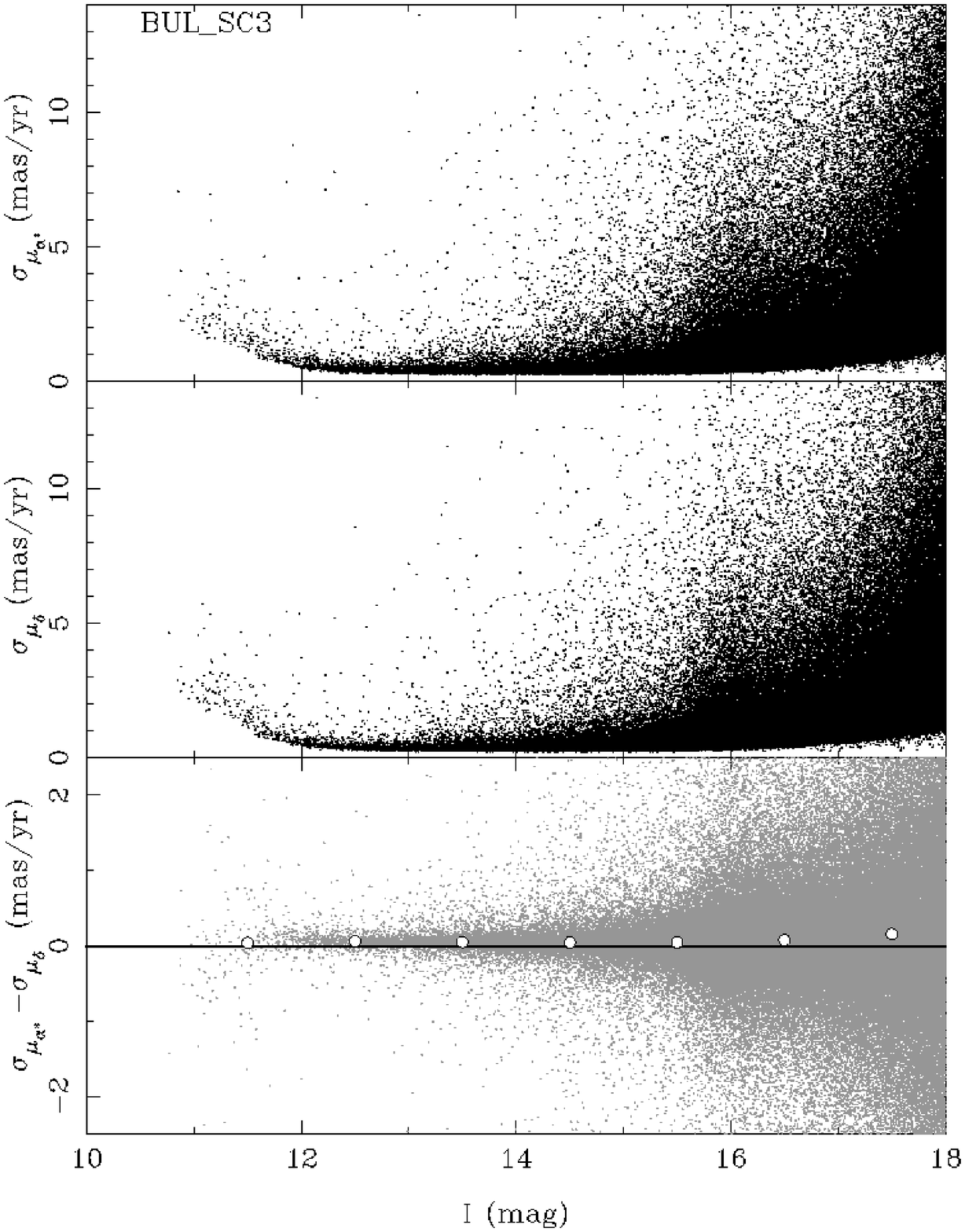}
\caption{Uncertainties in $\mu_{\alpha*}$ ($\sigma_{\mu_{\alpha*}}$, upper panel), 
$\mu_\delta$ ($\sigma_{\mu_\delta}$, middle panel) and the 
difference between them ($\sigma_{\mu_{\alpha*}} - \sigma_{\mu_\delta}$, lower panel) 
in OGLE-II field BUL\_SC3 as a function of the $I$-band magnitude.  
In the lower panel, where open circles represent mean values 
for each 1 magnitude bin, we can see that $\sigma_{\mu_{\alpha*}}$ is 
systematically larger than $\sigma_{\mu_\delta}$ by $\sim 0.1$ mas\,yr$^{-1}$ level.
  \label{fig:sigma}
  }
\end{center}
\end{figure}

In Fig. \ref{fig:difref} we present a colour -- magnitude diagram (CMD) 
for stars in OGLE field BUL\_SC5, which is one of the most  reddened OGLE-II fields. 
We also show the correlation between the differential refraction coefficient
$a$ and the apparent $V-I$ colour for stars with $I<16$ in this field. 
This correlation for other fields is similar, but it has a slight
dependence on the population of the majority of stars in each field.
In Fig. \ref{fig:sigma}, we plot the uncertainties in 
$\mu_{\alpha*}$ ($\sigma_{\mu_{\alpha*}}$, upper panel), 
in $\mu_\delta$ ($\sigma_{\mu_\delta}$, middle panel) and  
difference between them ($\sigma_{\mu_{\alpha*}} - \sigma_{\mu_\delta}$, lower panel)  
in BUL\_SC3 as a function of the apparent $I$-band magnitude.
In the lower panel, where open circles represent mean values  
for each 1 magnitude bin, we can see that $\sigma_{\mu_{\alpha*}}$ is systematically 
larger than $\sigma_{\mu_\delta}$ at $\sim 0.1$ mas\,yr$^{-1}$ level.
We see the same trend in all our fields. This trend is expected from 
the residual scattering due to the systematic correction and the 
differential refraction which is large in 
the direction of $\alpha$.  The mean uncertainties 
$\langle\sigma_\mu\rangle\equiv \langle\sqrt{\sigma_{\mu_{\alpha*}}^2 
+\sigma_{\mu_\delta}^2}\rangle$ with 2$\sigma$ clipping 
as a function of $I$ are listed in Table \ref{tbl:fields}.  
Note: that the accuracy of proper motions in our
catalogue is better than 1 mas\,yr$^{-1}$ for $12<I<14$.

\begin{figure}
\begin{center}
\includegraphics[angle=-90,scale=0.35,keepaspectratio]{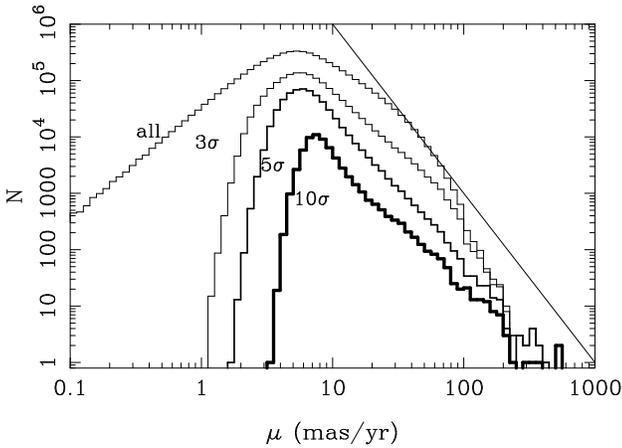}
\caption{Histograms are shown for $\mu$ for all $N = 5,080,236$ stars, and
for measurements with confidence better than $3\sigma$, $5\sigma$ and
$10\sigma$ (from the thinnest to the thickest line, respectively) for all
49 OGLE-II fields. The total number of
stars with proper motion with $3\sigma$, $5\sigma$ and $10\sigma$ accuracy
are $N_3 = 1,469,838$, $N_5=568,665$, $N_{10}=61,231$.
The inclined solid line: ($log(N)=-3log(\mu)+const.$) has the slope expected
from stars that have uniform distribution and kinematics.
\label{fig:hpmhist}
}
\end{center}
\end{figure}

Fig. \ref{fig:hpmhist} shows the histogram of our proper motion
measurements $\mu$ for the whole catalogue, with the lines with increasing
thickness corresponding to all stars, and to those with the proper motion
detected with confidence better than $3\sigma$, $5\sigma$ and $10\sigma$,
respectively.  The total number of stars with proper motions measured
to $3\sigma$, $5\sigma$ and $10\sigma$ accuracy
is $N_3 = 1,469,838$, $N_5=568,665$, $N_{10}=61,231$, respectively.
The inclined solid line: ($log(N)=-3log(\mu)+const.$), has a slope corresponding
to the expectation for a uniform distribution and kinematics of stars in space.
The distribution of accurate ($5\sigma$ and $10\sigma$) proper motions seems
to be roughly consistent with the uniform distribution.

However, the number of very high proper motion stars,
($\mu>200$ mas\,yr$^{-1}$), appears to be smaller than
expected from a uniform distribution. This may be
due to saturation of images of nearby, and therefore apparently
bright, stars in OGLE images.

The distribution of less accurate ($\le 3\sigma$) proper motions
has an apparent cut-off at $\mu \sim 100$ mas\,yr$^{-1}$.
Such stars are apparently faint, their positions
have large errors, and they may be difficult to identify
in a search radius of 1 pixel,
which corresponds to  $\mu \sim 100$ mas\,yr$^{-1}$.

A more thorough analysis of various selection effects is beyond
the scope of this paper.  Readers are advised to use caution
in a statistical analysis of our catalogue.

\section{Zero point}
\label{sec:zeropoint}
The proper motions in our catalogue are relative values. 
We need QSOs behind our fields to define the 
zero point of our proper motions in the inertial frame.
However we can get rough information of the zero point of 
our proper motions by measuring the average proper motion of
stars located in the GB, which is presumably close to absolute 
proper motion of the Galactic Center (GC).

\begin{figure}
\begin{center}
\includegraphics[angle=-90,scale=0.36,keepaspectratio]{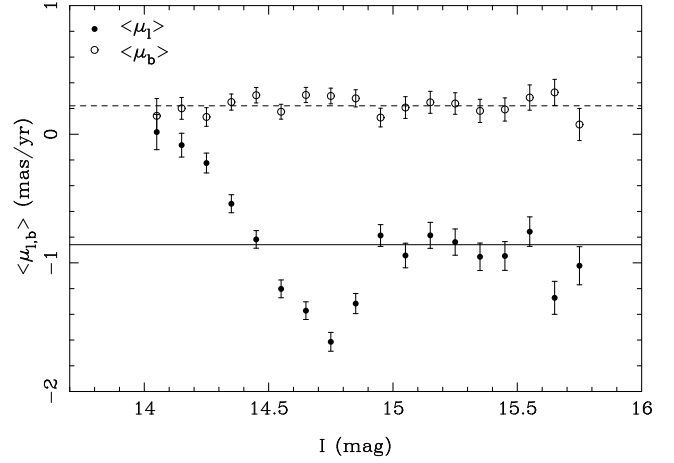}
\caption{Mean proper motions of stars at GC $\langle \mu_l \rangle$ 
(filled circle) and $\langle \mu_b \rangle$ (open circle) within 
$\Delta I =0.1$ mag bin as a function of $I$-band magnitude. Solid and
dashed lines represent the mean of these bins without weighting by errors.
We can see the streaming motion of bright ($I\sim 14.1$) and faint 
($\sim 14.7$) RCGs. The proper motion of Red Giants ($I>15$) is the mid 
of them as expected from that they are on average on the GC.
  \label{fig:muGC}
}
\end{center}
\end{figure}

We select stars in the GB defined in Fig. \ref{fig:cmd}. 
Here we choose only objects whose proper
motion accuracy is better than 2.5 mas\,yr$^{-1}$. 
We divide them into bins with a width of $\Delta I =0.1$ mag in $I$-band 
magnitude and take mean of their proper motions for each bin. 
We plot these mean proper motions $\langle \mu_{l,b} \rangle$ 
as a function of $I$ for BUL\_SC2 in Fig. \ref{fig:muGC}.
In this figure we can see the streaming motion of bright ($I\sim 14.1$)
and faint ($I\sim 14.7$) RCGs in $\mu_l$. The proper motion of Red Giants
($I>15$) is the same as the average motion of the RCGs, as expected since 
Red Giants are on average in the GC.

To measure the proper motion of the GC, $\mu_{l, \rm GC}$ and
 $\mu_{b, \rm GC}$ in our reference frame, without any bias due to 
the incompleteness for fainter stars,
we take a mean of these $\langle \mu_{l,b} \rangle$ without 
weighting by their errors. 
This provides an estimate that is more reliable than taking
the mean proper motion of all individual stars.
The measured $\mu_{l, \rm GC}$ and  $\mu_{b, \rm GC}$
are shown as a solid and dashed line for BUL\_SC2 in Fig. \ref{fig:muGC} 
and listed for other fields in Table \ref{tbl:muGC} along with those 
in equatorial coordinates $(\mu_{\alpha*, \rm GC}, \mu_{\delta, \rm GC})$.
Table \ref{tbl:muGC} is also available in electronic format via 
anonymous ftp with the main catalogue (see \S\,\ref{sec:catalog}).

The proper motion measurements in our catalogue can be transformed to
the inertial frame by formula;

\begin{eqnarray}
 \label{eq:pma_inertial}
\mu_{\alpha* \rm OGLE} &=& \mu_{\alpha*} - \mu_{\alpha*, \rm GC} + \mu_{\alpha*, \rm GC, inert} \\
\mu_{\delta \rm OGLE} &=& \mu_{\delta} - \mu_{\delta, \rm GC} + \mu_{\delta, \rm GC, inert}.
 \label{eq:pmd_inertial}
\end{eqnarray}
Here
$(\mu_{\alpha*, \rm GC, inert}, \mu_{\delta, \rm GC, inert})= (-2.93, -5.17)$ 
mas\,yr$^{-1}$ is the expected proper motion of the GC
relative to the inertial frame, 
assuming a flat rotation curve of $v_r \sim 220$ km\,s$^{-1}$, the distance 
between the GC and the Sun of $R_0 =8.0$ kpc (\citealt{eis03}) and Solar
velocity of $(v_{\odot l},v_{\odot b})$= (5.25 km\,s$^{-1}$, 7.17 km\,s$^{-1}$)
relative to the Local Standard of Rest (RSL) (\citealt{deh98}).
This transformation gives us crude absolute proper motions, 
while this gives us very good relative zero points from field to field
as we show later.

The reader can apply this transformation formula to our catalogue
to get value in the inertial frame. 
We didn't apply this transformation to our catalog because
the transformation to the inertial frame can be improved 
with the QSOs to be discovered behind our fields in the future.

\begin{table*}
\caption{Proper motions of the GC for all 49 OGLE-II fields. 
The number of stars used in measurements $N_{star}$, proper motions of the GC
in our reference frame in equatorial coordinate $\mu_{\alpha,\rm GC}$ and 
$\mu_{\delta,\rm GC}$ and galactic coordinate  $\mu_{l,\rm GC}$ and $\mu_{b,\rm GC}$
with their errors. 
\label{tbl:muGC}}
\begin{tabular}{lccccccccc}\\
field&  $N_{star}$ 
&  $\mu_{\alpha,\rm GC}$ & $\sigma_{\mu_{\alpha,\rm GC}}$& $\mu_{\delta,\rm GC}$ & $\sigma_{\mu_{\delta,\rm GC}}$ 
&  $\mu_{l,\rm GC}$ & $\sigma_{\mu_{l,\rm GC}}$& $\mu_{b,\rm GC}$ & $\sigma_{\mu_{b,\rm GC}}$ \\
& &   \multicolumn{8}{c}{(mas\,yr$^{-1}$)} \\
BUL\_SC1 &  23938 & -0.57 &  0.07 & -0.55 &  0.11  & -0.76 &  0.12 &  0.23 &  0.04\\
BUL\_SC2 &  25738 & -0.61 &  0.05 & -0.64 &  0.09  & -0.86 &  0.10 &  0.22 &  0.02\\
BUL\_SC3 &  43597 & -0.66 &  0.05 & -0.68 &  0.08  & -0.92 &  0.09 &  0.22 &  0.02\\
BUL\_SC4 &  42847 & -0.57 &  0.06 & -0.62 &  0.10  & -0.82 &  0.11 &  0.18 &  0.04\\
BUL\_SC5 &  12704 & -0.72 &  0.08 & -0.94 &  0.12  & -1.18 &  0.14 &  0.14 &  0.05\\
BUL\_SC6 &  10167 & -0.61 &  0.05 & -0.80 &  0.08  & -0.99 &  0.09 &  0.15 &  0.03\\
BUL\_SC7 &   9431 & -0.55 &  0.06 & -0.68 &  0.10  & -0.86 &  0.11 &  0.16 &  0.03\\
BUL\_SC8 &   7397 & -0.98 &  0.05 & -0.97 &  0.08  & -1.31 &  0.09 &  0.41 &  0.04\\
BUL\_SC9 &   7364 & -0.85 &  0.05 & -1.02 &  0.08  & -1.30 &  0.09 &  0.28 &  0.03\\
BUL\_SC10 &   8476 & -0.84 &  0.06 & -0.98 &  0.11  & -1.26 &  0.12 &  0.28 &  0.03\\
BUL\_SC11 &   8067 & -0.80 &  0.06 & -1.03 &  0.10  & -1.28 &  0.11 &  0.23 &  0.03\\
BUL\_SC12 &   9913 & -0.95 &  0.05 & -1.33 &  0.08  & -1.62 &  0.09 &  0.20 &  0.03\\
BUL\_SC13 &   9899 & -0.84 &  0.06 & -1.13 &  0.10  & -1.39 &  0.11 &  0.20 &  0.03\\
BUL\_SC14 &  19398 & -0.57 &  0.05 & -0.70 &  0.08  & -0.89 &  0.10 &  0.12 &  0.01\\
BUL\_SC15 &  16454 & -0.53 &  0.05 & -0.79 &  0.08  & -0.95 &  0.09 &  0.04 &  0.03\\
BUL\_SC16 &  15801 & -0.71 &  0.06 & -1.00 &  0.09  & -1.22 &  0.10 &  0.15 &  0.04\\
BUL\_SC17 &  16665 & -0.68 &  0.05 & -0.97 &  0.07  & -1.18 &  0.08 &  0.13 &  0.04\\
BUL\_SC18 &  23064 & -0.60 &  0.05 & -0.73 &  0.09  & -0.93 &  0.10 &  0.17 &  0.02\\
BUL\_SC19 &  21441 & -0.60 &  0.05 & -0.77 &  0.08  & -0.96 &  0.09 &  0.16 &  0.02\\
BUL\_SC20 &  31933 & -0.42 &  0.06 & -0.61 &  0.10  & -0.74 &  0.11 &  0.06 &  0.04\\
BUL\_SC21 &  30486 & -0.55 &  0.06 & -0.71 &  0.09  & -0.89 &  0.11 &  0.13 &  0.03\\
BUL\_SC22 &  31278 & -0.57 &  0.05 & -0.60 &  0.08  & -0.80 &  0.10 &  0.20 &  0.02\\
BUL\_SC23 &  27679 & -0.59 &  0.06 & -0.70 &  0.10  & -0.91 &  0.11 &  0.16 &  0.02\\
BUL\_SC24 &  27355 & -0.55 &  0.06 & -0.59 &  0.08  & -0.79 &  0.09 &  0.18 &  0.03\\
BUL\_SC25 &  25782 & -0.52 &  0.05 & -0.66 &  0.09  & -0.83 &  0.10 &  0.12 &  0.02\\
BUL\_SC26 &  21577 & -0.62 &  0.05 & -0.66 &  0.07  & -0.88 &  0.08 &  0.19 &  0.02\\
BUL\_SC27 &  20250 & -0.60 &  0.04 & -0.72 &  0.07  & -0.92 &  0.08 &  0.15 &  0.02\\
BUL\_SC28 &  12159 & -0.73 &  0.05 & -0.78 &  0.07  & -1.04 &  0.07 &  0.22 &  0.04\\
BUL\_SC29 &  11755 & -0.75 &  0.04 & -0.78 &  0.05  & -1.06 &  0.06 &  0.24 &  0.02\\
BUL\_SC30 &  29140 & -0.54 &  0.05 & -0.64 &  0.08  & -0.82 &  0.10 &  0.15 &  0.03\\
BUL\_SC31 &  28389 & -0.44 &  0.06 & -0.67 &  0.10  & -0.80 &  0.11 &  0.06 &  0.03\\
BUL\_SC32 &  25750 & -0.53 &  0.06 & -0.65 &  0.09  & -0.82 &  0.10 &  0.15 &  0.03\\
BUL\_SC33 &  23709 & -0.61 &  0.06 & -0.69 &  0.10  & -0.90 &  0.12 &  0.20 &  0.02\\
BUL\_SC34 &  32041 & -0.53 &  0.06 & -0.59 &  0.09  & -0.78 &  0.11 &  0.17 &  0.03\\
BUL\_SC35 &  25997 & -0.56 &  0.05 & -0.66 &  0.09  & -0.85 &  0.10 &  0.17 &  0.03\\
BUL\_SC36 &  24278 & -0.59 &  0.05 & -0.64 &  0.08  & -0.85 &  0.10 &  0.21 &  0.02\\
BUL\_SC37 &  38711 & -0.68 &  0.06 & -0.75 &  0.09  & -0.99 &  0.10 &  0.20 &  0.03\\
BUL\_SC38 &  25633 & -0.46 &  0.06 & -0.64 &  0.10  & -0.79 &  0.12 &  0.08 &  0.03\\
BUL\_SC39 &  41129 & -0.47 &  0.06 & -0.64 &  0.11  & -0.79 &  0.12 &  0.08 &  0.02\\
BUL\_SC40 &  25729 & -0.52 &  0.05 & -0.57 &  0.08  & -0.76 &  0.10 &  0.16 &  0.02\\
BUL\_SC41 &  25665 & -0.50 &  0.05 & -0.56 &  0.07  & -0.74 &  0.08 &  0.15 &  0.03\\
BUL\_SC42 &  18649 & -0.79 &  0.06 & -0.88 &  0.10  & -1.15 &  0.11 &  0.26 &  0.04\\
BUL\_SC43 &  25012 & -0.36 &  0.07 & -0.36 &  0.11  & -0.50 &  0.13 &  0.11 &  0.03\\
BUL\_SC44 &     -- &  --   &  --   &  --   &  --    &  --   &  --   &  --   &  --  \\
BUL\_SC45 &  13776 & -0.43 &  0.08 & -0.32 &  0.14  & -0.49 &  0.15 &  0.22 &  0.03\\
BUL\_SC46 &  12608 & -0.40 &  0.08 & -0.23 &  0.14  & -0.40 &  0.16 &  0.24 &  0.04\\
BUL\_SC47 &   6294 & -1.18 &  0.07 & -1.20 &  0.10  & -1.65 &  0.12 &  0.31 &  0.03\\
BUL\_SC48 &   6965 & -1.09 &  0.06 & -1.00 &  0.08  & -1.44 &  0.10 &  0.35 &  0.03\\
BUL\_SC49 &   6299 & -0.99 &  0.06 & -0.86 &  0.08  & -1.26 &  0.09 &  0.35 &  0.04\\
\end{tabular}
\end{table*}

\begin{figure}
\begin{center}
\includegraphics[angle=0,scale=0.5,keepaspectratio]{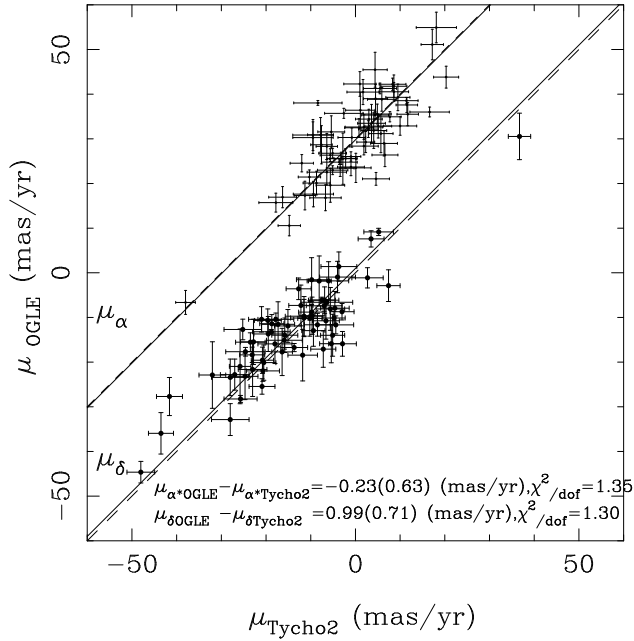}
\caption{Comparison of proper motions $\mu_\alpha*$ (thin) and $\mu_\delta$ 
(thick) for 65 stars cross-identified in our catalogue and in Tycho-2 catalogue.
The proper motions in our catalogue are transformed to
the inertial frame by equations (\ref{eq:pma_inertial}) and (\ref{eq:pmd_inertial}). 
Dashed lines indicate $\mu_{OGLE}=\mu_{Tycho-2}$, and solid lines
represent best fit with the offsets:
$\mu_{\alpha* \rm OGLE} - \mu_{\alpha* \rm Tycho-2} =-0.23$ mas\,yr$^{-1}$ and 
$\mu_{\delta \rm OGLE}  - \mu_{\delta \rm Tycho-2}  =0.99$ mas\,yr$^{-1}$.
$\mu_\alpha*$ are shifted for +30 mas\,yr$^{-1}$ for clarity.
  \label{fig:crossref_tycho}
}
\end{center}
\end{figure}

To check our measurements we cross-identified stars in our catalogue
with the Tycho-2 catalogue (\citealt{hog00}).  We selected from our
catalogue objects with proper motions higher than 10  mas\,yr$^{-1}$
and measured with a significance above 3$\sigma$,
to avoid mis-identification.
Most high proper motion stars in Tycho-2 are saturated in OGLE images.
We found 65 Tycho-2 stars in our catalogue and their proper motions: 
$\mu_{\alpha*}$ (thin) and $\mu_\delta$ (thick) are presented in 
Fig. \ref{fig:crossref_tycho}. 
Here OGLE proper motions have been transformed into the inertial frame 
by equations (\ref{eq:pma_inertial}) and (\ref{eq:pmd_inertial}).
We can see a very good correlation between OGLE and Tycho-2 measurements. 
Dashed lines indicate $\mu_{OGLE}=\mu_{Tycho-2}$, and solid lines
represent best fit with fixed unit slope and a possible offset. 
A good correspondence between Tycho-2 and OGLE-II  proper motions
gives us certain confidence in our measurements. 
Slightly larger offsets in $\mu_\delta$ imply that the error in absolute 
proper motion is at a level of 1 mas\,yr$^{-1}$, which 
can be improved by using QSOs in the near future. 

\begin{figure}
\begin{center}
\includegraphics[angle=-90,scale=0.5,keepaspectratio]{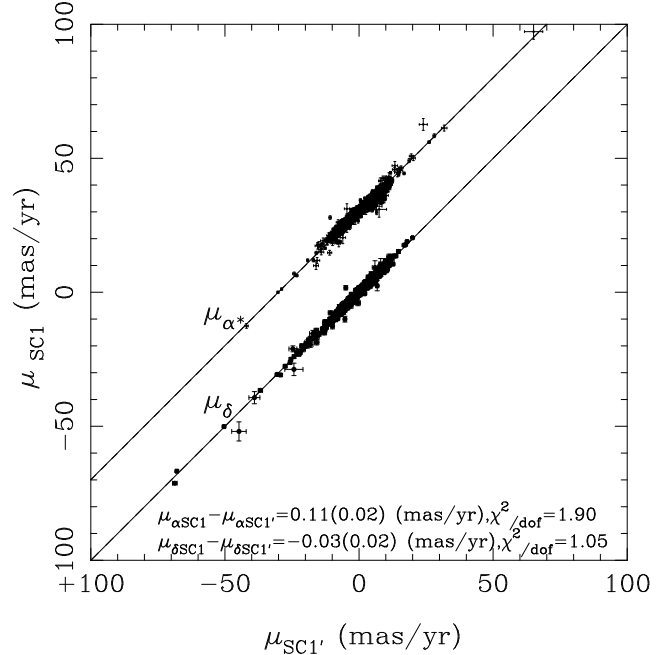}
\caption{Same figure as Fig. \ref{fig:crossref_tycho} for
1368 stars cross-identified in BUL\_SC1 by \citealt{sum03a} (hereafter SC1$'$) 
and in SC1 by this work, measured with better than $10\sigma$ level accuracy.
  Dashed lines indicate $\mu_{SC1}=\mu_{SC1'}$ which overlap with the solid line 
  in $\mu_\delta$, and solid lines represent the best fit with the offsets:
  $\mu_{\alpha* SC1}-\mu_{\alpha* SC1'}=0.11$ mas\,yr$^{-1}$  and
  $\mu_{\delta SC1}-\mu_{\delta SC1'}=-0.03$  mas\,yr$^{-1}$.
  $\mu_\alpha$ are shifted for +30 mas\,yr$^{-1}$ for clarity. Note: 
  the scale is different from  Fig. \ref{fig:crossref_tycho}.
  \label{fig:crossref_sc1old-1}
}
\end{center}
\end{figure}

We also compared the measurements in BUL\_SC1 made by \citealt{sum03a} 
(hereafter SC1$'$) with the proper motions presented in this paper,
for which our measurements are above the $10\sigma$ level of accuracy.
The two sets of proper motions for 1368 cross-identified stars are shown in
Fig. \ref{fig:crossref_sc1old-1} together with the best fit line.  The
offset between the two sets, and the differences between individual
measurements are within estimated errors.
Note: the scale is different than in Fig. \ref{fig:crossref_tycho}.
Large reduced chi square in $\mu_\alpha$ are because \citealt{sum03a}
didn't correct systematic distortions (see \S \ref{sec:analysis}),
though their systematic distortions have been reduced at a level of 
2 mas\,yr$^{-1}$ by dividing images into small chunks.

\begin{figure}
\begin{center}
\includegraphics[angle=0,scale=0.5,keepaspectratio]{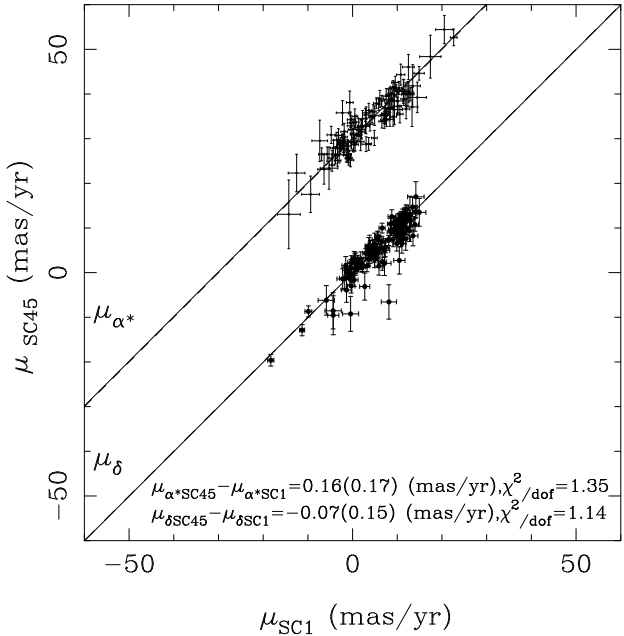}
\caption{Same figure as Fig. \ref{fig:crossref_tycho} for
115 stars cross-identified in the overlap region between OGLE-II fields
BUL\_SC1 and BUL\_SC45, and measured with better than $5\sigma$ level accuracy.
Proper motions are transformed into the value in the inertial frame by 
equations (\ref{eq:pma_inertial}) and (\ref{eq:pmd_inertial}).
Dashed lines indicate $\mu_{SC1}=\mu_{SC45}$, and solid lines
represent the best fit with the offsets:
  $\mu_{\alpha* SC1}-\mu_{\alpha* SC45}=0.16$ mas\,yr$^{-1}$  and
  $\mu_{\delta SC1}-\mu_{\delta SC45}=-0.07$  mas\,yr$^{-1}$.
  $\mu_\alpha$ in SC45 are shifted for +30 mas\,yr$^{-1}$ for clarity.
  \label{fig:crossref_sc1-45}
}
\end{center}
\end{figure}

We compared the measurements done by us in one of the overlap regions: 
between fields BUL\_SC1 and BUL\_SC45.  The proper motions of 115
cross-identified stars, and the best fit with a small offset between
zero points, are presented in Fig. \ref{fig:crossref_sc1-45}.
Here proper motions have been transformed into the inertial frame
by equations (\ref{eq:pma_inertial}) and (\ref{eq:pmd_inertial}). 
Good correlations between  them with a rather small zero point offsets: 
$\mu_{\alpha* SC1}-\mu_{\alpha* SC45}=0.16$ mas\,yr$^{-1}$ and
$\mu_{\delta SC1}-\mu_{\delta SC45}=-0.07$  mas\,yr$^{-1}$
gives us certain confidence in our measurements and in equations
(\ref{eq:pma_inertial}) and (\ref{eq:pmd_inertial}) in terms of the 
relative the zero point.  The scatter is also consistent with the 
estimated errors.

\section{Possible problems}
\label{sec:problem}
There can be various problems associated with proper motions of variable stars.    
Our fields are very crowded, hence many stars may be blended.
In the case of blending we measure an average position of a blend of
several stars within a seeing disk.
If a variable star is blended with other stars which have
slightly different positions within a seeing disk, the average position
of the blended image may change while one component of the blend varies.
This change might mimic a proper motion.  

\begin{figure}
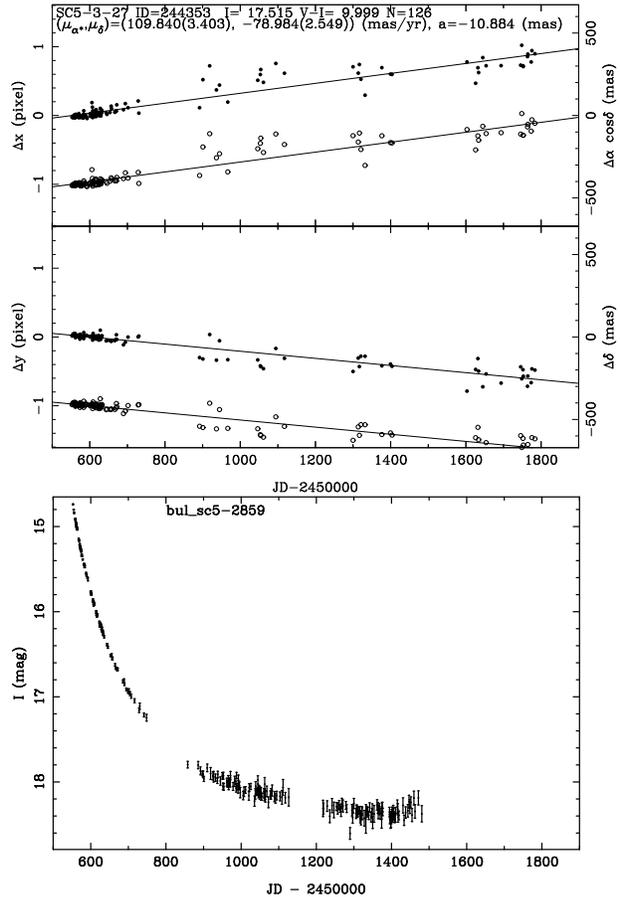

\includegraphics[angle=-90,scale=0.35,keepaspectratio]{fig15a.eps}
\includegraphics[angle=-90,scale=0.32,keepaspectratio]{fig15b.eps}
\begin{center}
\caption{Top panel: Same plots as Fig. \ref{fig:shiftcurve} for the star ID=244353 
in the OGLE-II field BUL\_SC5. Filled circles represent actual positions and
open circles are correspond to positions corrected for
differential refraction, with the offset of $-1$ pixels.
Bottom panel: $I$-band light curve of same star (ID=2859 in catalogue of 
\citealt{woz01}), a possible very long microlensing event.
\label{fig:pmmicrolensing}
}
\end{center}
\end{figure}

As an example we present the time variation of the position of a very long 
timescale microlensing event candidate (\citealt{smi03})
in the top panel of Fig. \ref{fig:pmmicrolensing}.   
The $I$-band light curve of this event is shown in the 
bottom panel of Fig.  \ref{fig:pmmicrolensing} (ID=2859 in variable 
star catalogue of \citealt{woz02}).
In the top panel we can see an apparent proper motion in the first year which 
is coincident with the apparent brightness fading, as shown in the
bottom panel.  After the second year the proper motion
seems to be small, and the position data points are sparse, which coincides
with the low brightness of the star and may reflect the presence of a blend --
in poor seeing DoPHOT cannot resolve the two stars, hence there are very few data
points in the upper panel.
The blend was actually found in the higher resolution
OGLE-III images in 2002. 

A plausible interpretation is that the microlensed star was brighter
than the nearby faint `companion' star in 1997, but by 2000 it became
the fainter of the two.  The position of the `companion' is consistent 
with the direction of the apparent proper motion.
The faint stars seem to have low proper motion.
High resolution observations are needed to fully understand this object.

\begin{figure}
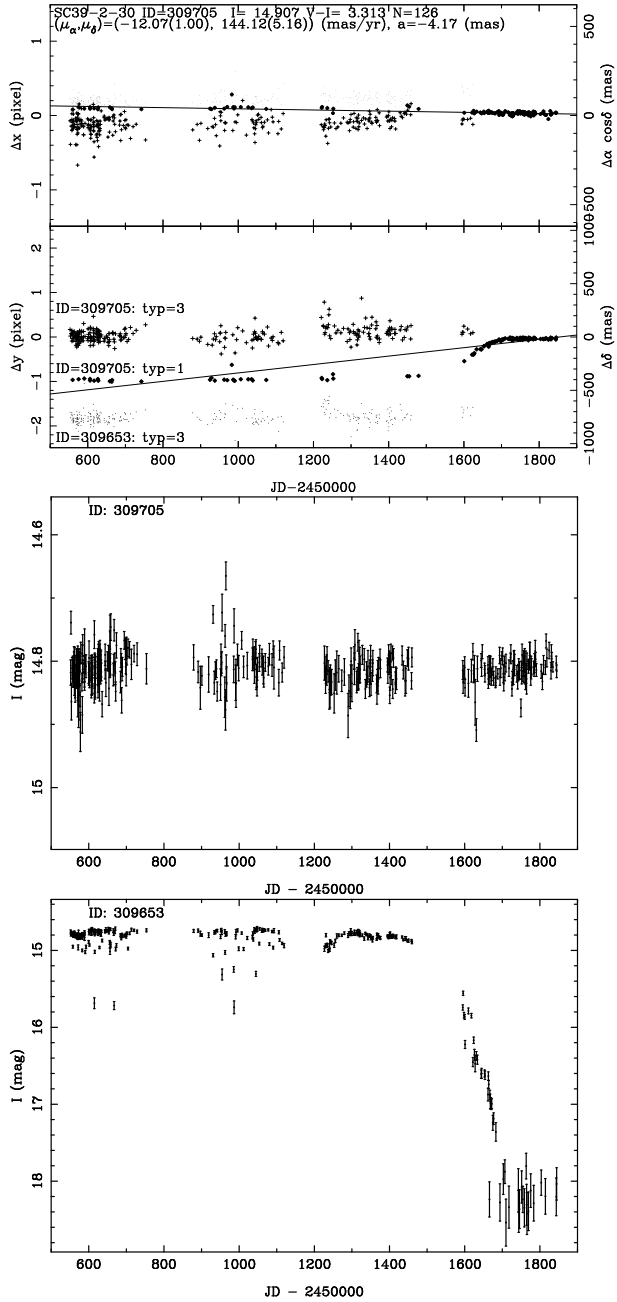

\includegraphics[angle=-90,scale=0.35,keepaspectratio]{fig16a.eps}
\includegraphics[angle=-90,scale=0.32,keepaspectratio]{fig16b.eps}
\includegraphics[angle=-90,scale=0.32,keepaspectratio]{fig16c.eps}
\begin{center}
\caption{
Top: Same plots as Fig. \ref{fig:shiftcurve} for star ID=309705 in OGLE-II field
BUL\_SC39. Filled circles represent actual positions with type=1 (used in this 
work), crosses correspond to position measurements with type=3 (not used)
and dots indicate positions measurements of neighboring star ID=309653 with type=3 
which are shifted by +0.18 pixels in x and -1.89 pixels in y,
i.e. these dots are as they are on CCD, relative to ID=309705. 
This is a probable case of a blend with one component being a variable star.
\label{fig:pmvariable}
}
\end{center}
\end{figure}

As another example we present in Fig. \ref{fig:pmvariable} the
time variation of the position of a star ID=309705 in the OGLE-II field
BUL\_SC39. 
Filled circles represent measured positions with type=1 (used in this work),
which were at a fixed location for the first three years but moved significantly
in the fourth season.  The star ID=309705 identified these centroids at 
the edge of the search radius (y=-1) for the first 3 seasons and at the 
center (y=0) in the 4th season.  On the other hand, the neighboring star 
ID=309653 which has a similar brightness of $I=14.8$ 
and position of 0.18 pixels East (positive x) and -1.89 pixels South (negative y), 
identified the same centroids at the edge of the search radius (y=-1).

To see the details of this object, in Fig. \ref{fig:pmvariable}
we also plot the position measurements which are categorized as 
a star blended with other stars (type=3, which are not used in this work) 
by DoPHOT for this star (crosses) and for a neighboring star ID=309653 (dots),
which are shifted by +0.18 pixels in x and -1.89 pixels in y,
i.e. these dots are as they are on the CCD, relative to ID=309705.
We can see that these positions (crosses and dots) are identified around
 y=0 (ID=309705) and -2 (ID=309653), respectively, with filled circles
in-between them. 
We also show the $I$-band light curves of ID=309705 (middle panel) and 
ID=309653 (bottom panel) in Fig. \ref{fig:pmvariable}. We can see ID=309705 
is constant during four seasons but ID=309653 suddenly faded during the 4th season.
This is likely to be a R CrB type variable.

The simplest interpretation is as follows; 
small number of data points in the seasons 1997-99 indicates that DoPHOT found 
that object which is composite of these two stars only on the bad seeing frames as a single
star with type=1, in other cases they are separated but categorized as blended - type=3. 
In 2000 when the star ID=309653 faded, the centroid of the composite  moved
and finally the star ID=309705 became a "single" star  with type=1.
So the number of data points is large in 2000.

Proper motions of variable stars may have their
errors increased not only because of variable contribution of blending, 
but also because variable stars may change colours, and therefore the
coefficient of differential refraction may also change.  In rare cases
of very long period variables this may be noticeable.  
Note: we do not treat this kind of objects in a special way, so
a reader must be careful when using our catalogue in studies of variable stars.

The effect of blending changes with seeing may contribute to the scatter
of data points, but it is not likely to have a seasonal effect.  Hence,
we think that seeing variations are not a major problem.

The probability of blending is much larger for fainter stars,
in particular those close to $I = 18$ mag.
This may produce a bias in their proper motions, most likely reducing
their formal proper motion, as the blended stars may have a different 
proper motion vector, so the average value is likely to be reduced. 
We also do not provide a special treatment for this kind of objects, so a
reader must be careful when using our catalogue for faint stars.


\section{Discussion and Conclusion}

\label{sec:disc}

We have measured proper motions for 5,080,236 stars in all 49 OGLE-II
GB fields, covering a range of $-11^\circ <l< 11^\circ$ and 
$-6^\circ <b<3^\circ$.  Our catalogue contains objects with proper 
motions up to $\mu = 500$ mas\,yr$^{-1}$
and $I$-band magnitudes in the range $11\le I\le 18$.
The accuracy of proper motions in our catalogue is better 
than 1 mas\,yr$^{-1}$ for $12<I<14$.

One should keep in mind that all measurements of $\mu$ presented here
are not absolute, but relative to the astrometric reference frame
which is roughly that of the Galactic Center (GC) with a small offset
seen in the lower panel of Fig. \ref{fig:muvsx}. However, as demonstrated
in \S \ref{sec:zeropoint}, by using the crude estimation for the proper
motion of the GC in our reference frame and formula equations 
(\ref{eq:pma_inertial}) and (\ref{eq:pmd_inertial}),
we can obtain crude proper motions in an inertial frame.
From the comparison with these inertial values and the Tycho2 catalogue, 
this transformation seems to work well with errors at a level of 1 mas\,yr$^{-1}$.
From comparison of $\mu$ measured in the overlap region of fields BUL\_SC1
and BUL\_SC45 (Fig. \ref{fig:crossref_sc1-45}), this transformation
works very well in the relative offset from field to field. 
These zero points for proper motions can be improved
by using background quasars which may be detected in the near future
using the OGLE-II variability catalogue (\citealt{woz02,eye02,dob03}).

As demonstrated by \cite{sum03a}, the proper motions
based on OGLE-II data can be 
used to clearly detect the presence of a strong streaming motion
(rotation) of stars in the Galactic bar. 
While the reference frame established from 
all stars is not well defined with respect to the inertial
frame, the relative motions of groups of stars within
a given field are well determined.  

Though our primary goal is to constrain the Galactic bar model with
the future analysis of our catalogue, we provide proper motions
for all stars with $I<18$ mag in all 49 OGLE-II GB fields
because this catalogue might be useful for a variety of projects.
An analysis of the catalogue is beyond the scope
of the present study.

\section*{Acknowledgments}

We are grateful to B. Paczy\'{n}ski for helpful comments and
discussions.  We acknowledge M. Smith for carefully reading the 
manuscript and helpful comments.
We are also thankful to the referee, F. van Leeuwen for 
suggestive comments.
T.S.  acknowledge the financial support from the
Nishina Memorial Foundation and JSPS.
The paper was partly supported by the Polish KBN grant  2P03D02124 to
A.\ Udalski.
This work was partly supported with the following grants to 
B. Paczy\'nski:
NSF grants AST-9820314 and AST-0204908, and NASA grants NAG5-12212,
and grant HST-AR-09518.01A provided by NASA through a grant from
the Space Telescope Science Institute, which is operated by the Association of
Universities for Research in Astronomy, Inc., under NASA contract NAS5-26555.

\label{lastpage}
\clearpage

\end{document}